\documentclass[showpacs,preprintnumbers,nofootinbib,amsmath,amssymb,aps,prd,floatfix]{revtex4-1}
\pdfoutput=1
\usepackage{graphicx}
\usepackage{bm}
\usepackage{subfigure}
\usepackage{url}
\usepackage[hyperindex,colorlinks=true,linkcolor = blue,
anchorcolor = blue,
citecolor = darkgreen,
filecolor = blue,
urlcolor = blue]{hyperref}
\usepackage{color}
\usepackage{bigdelim}
\usepackage{booktabs}
\usepackage{multirow}
\usepackage{cancel}
\usepackage{stackrel}
\usepackage{paralist}
\usepackage{xspace}
\usepackage{catchfile}
\usepackage{enumitem}
\usepackage[normalem]{ulem}
\usepackage[utf8]{inputenc}
\usepackage[english]{babel}

\usepackage{float}
\usepackage{multirow}

\definecolor{darkgreen}{rgb}{0.0, 0.5, 0.0}

\makeatletter
\def\l@subsubsection#1#2{}
\makeatother

\def\kpc{{\rm\,kpc}}

\def\msun{{\rm\,M_\odot}}

\def\GeVcm{{\rm \, GeV/cm^3}}


\newcommand{\localDM}{\rho_{{\rm DM},\odot}}

\newcommand{\tabSpace}{0.15cm}
\newcommand{\cmrelation}{c_{200}\text{-}M_{200}}
\newcommand{\gNFW}{{\rm gNFW}}

\makeatletter

\renewcommand*{\p@subsection}{}

\renewcommand*{\p@subsubsection}{}
\makeatother

\begin{document}

\title{On the estimation of the Local Dark Matter Density using the rotation curve of the Milky Way}

\author{P.F.\ de Salas}\email{pablo.fernandez@fysik.su.se}
\affiliation{The  Oskar  Klein  Centre  for  Cosmoparticle  Physics,  Department  of Physics,  Stockholm  University, AlbaNova, 10691 Stockholm, Sweden}

\author{K. Malhan}
\affiliation{The  Oskar  Klein  Centre  for  Cosmoparticle  Physics,  Department  of Physics,  Stockholm  University, AlbaNova, 10691 Stockholm, Sweden}

\author{K. Freese}
\affiliation{The  Oskar  Klein  Centre  for  Cosmoparticle  Physics,  Department  of Physics,  Stockholm  University, AlbaNova, 10691 Stockholm, Sweden}
\affiliation{Nordita, KTH  Royal  Institute  of  Technology  and  Stockholm  University,  Roslagstullsbacken  23, 10691 Stockholm, Sweden}
\affiliation{Leinweber Center for Theoretical Physics, Department of
Physics, University of Michigan, Ann Arbor, MI 48109, USA}

\author{K. Hattori}
\affiliation{Department of Astronomy, University of Michigan, 1085 S. University Avenue, Ann Arbor, MI 48109, USA}

\author{M. Valluri}
\affiliation{Department of Astronomy, University of Michigan, 1085 S. University Avenue, Ann Arbor, MI 48109, USA}

\begin{abstract}

The rotation curve of the Milky Way is commonly used to estimate the local dark matter density $\localDM$. However, the estimates are subject to the choice of the distribution of baryons needed in this type of studies. In this work we explore several Galactic mass models that differ in the distribution of baryons and dark matter, in order to determine $\localDM$. For this purpose we analyze the precise circular velocity curve measurement of the Milky Way up to $\sim 25\kpc$ from the Galactic centre obtained from Gaia DR2 \citep{Eilers:1810.09466}. 
We find that the estimated value of $\localDM$ stays robust to reasonable changes in the spherical dark matter halo. However, we show that $\localDM$ is affected by the choice of the model for the underlying baryonic components. In particular, we find that $\localDM$ is mostly sensitive to uncertainties in the disk components of the Galaxy.
We also show that, when choosing one particular baryonic model, the estimate of $\localDM$ has an uncertainty of only about $10\%$ of its best-fit value, but this uncertainty gets much bigger when we also consider the variation of the baryonic model.
In particular, the rotation curve method does not allow to exclude the presence of an additional very thin component,
that can increase $\localDM$ by more than a factor of 8 
(the thin disk could even be made of dark matter). 
Therefore, we conclude that exclusively using the rotation curve of the Galaxy is not enough to provide a robust estimate of $\localDM$. 
For all the models that we study without the presence of an additional thin component, our resulting estimates of the local dark matter density take values in the range $\localDM \simeq \text{0.3--0.4}\GeVcm$, consistent with many of the estimates in the literature.

\end{abstract}


\maketitle

\tableofcontents
\newpage


\section{Introduction}\label{sec:intro}

An accurate and robust estimate of the local dark matter density $(\localDM)$ at the location of the Sun has been a goal in astrophysics for a long time.
A correct value is imperative for the interpretations of dark matter signals in a variety of experimental efforts.  For example, in direct detection experiments searching for Weakly Interacting Massive Particle (WIMPs), the parameter $\localDM$ is inversely proportional to the derived scattering cross section of the WIMP/nucleus interaction. Furthermore, since $\localDM$ is also linked to the normalization of the dark matter halo, a precise measurement of this quantity plays an important role for indirect detection searches, as well as in cosmology and astrophysics for those studies that rely on the dark matter distribution of the Milky Way.

One possible way of constraining $\localDM$ comes from the analysis of the rotation curve of the Milky Way. 
This method is based on the comparison of 
the observationally determined circular velocity (obtained from the measured rotation curve of our Galaxy\footnote{We note that the rotation curve and the circular velocity curve are often confused. 
The rotation curve is an observable quantity representing the mean rotational velocity at different radii. 
The measurement of the rotation curve of a galaxy is an essential part of determining the circular velocity curve, and this field is pioneered by the work of Vera Rubin~\cite{Rubin:1980zd} for external galaxies.}
with the radial Jeans equation)
and its theoretical estimates at different radii, $v_{\rm c} (R)$, 
where $R$ is the radius in the galactocentric cylindrical coordinates. The theoretical estimates of the circular velocity are derived from the potential $\Phi$ of the Galaxy through the relation
\begin{equation}\label{eq:vc-from-potential}
v_{\rm c}^2 (R) = R \left.\frac{\partial \Phi}{\partial R}\right|_{z = 0}.
\end{equation} 
The potential $\Phi$ is the sum of the contribution of different components $\Phi_i$. In our study we consider a spherical dark matter halo, a spherical Galactic bulge and axisymmetric distributions for the Galactic disk. In those cases in which a distribution is given in terms of its energy density $\rho_i$, the potential is obtained through the Poisson equation
\begin{equation}
\nabla^2 \Phi_i = 4\pi G \rho_i,
\end{equation}
where $G$ is the gravitational constant.
For a comprehensive review of the rotation curve's method and other methods to infer $\localDM$ we refer the reader to \cite{Read:2014qva}. 

Given our location in the Galaxy, determining the Milky Way's rotation curve to a good precision has been a persistent challenge \citep{Sofue:2011kw, Russeil:2017aa}. This has been partly due to the complexities that come into play from the dense and obscuring inter-galactic gas and dust present in the Galaxy's disk. However, the major drawback so far had been the lack of well measured 3D velocities and distances of the stars in the Milky Way. This handicap has now been overcome with the help of the second data release (DR2) of the ESA/Gaia mission \citep{GaiaDR2_2018_Brown, GaiaCollab2018kinematics}.

Gaia DR2 has provided extremely well measured proper motions and 2D positions on the sky for a significantly large volume of our Galaxy, as well as precise parallaxes within $\sim \text{2--3}\kpc$ from the solar neighbourhood. 
Combining this exquisite data with photometric information from 2MASS and WISE, and spectroscopic information from APOGEE, \cite{Hogg:1810.09468} estimated precise spectrophotometric parallaxes for red-giant stars up to $25\kpc$ from the Galactic centre.  
Using this information, 
\cite{Eilers:1810.09466} determined the Milky Way's rotation and circular velocity curves\footnote{Likewise based on Gaia DR2 data, \cite{Mroz:1810.02131} also measured the rotation curve of our Galaxy. Their result is consistent with that of \cite{Eilers:1810.09466}.} for the distance range of $5\kpc \leq R \leq 25\kpc$. 
Analyzing their accurate measurement of the rotation curve of the Galaxy, they also provide an estimate of $\localDM$.
In order to infer the value of $\localDM$ from the rotation curve, they use a Galactic mass model in which they fix the baryonic components and fit a spherical dark matter halo. With their assumptions, \cite{Eilers:1810.09466} quote a result of $\localDM = 0.30\pm0.03\GeVcm$.

In this paper we perform a more general study. We explore a variety of different Galactic mass models, comprised of different baryonic and dark matter components, to study the corresponding effects on the estimate of $\localDM$ from rotation curve measurements. Specifically, we reanalyze the circular velocity curve data from \cite{Eilers:1810.09466} to show that uncertainties in the baryonic components can lead to a significant uncertainty in the best-fit value of $\localDM$.

The paper is arranged as follows: In section~\ref{sec:models} we detail all the mass models that we use in our study; in section~\ref{sec:analyses} we present the analyses of the circular velocity curve based on different mass models; the results are compared and discussed in section~\ref{sec:discussion}, and we draw our conclusions in section~\ref{sec:conclusions}.

\section{Galactic models}\label{sec:models}

In order to study the influence that the choice of a specific mass model has in the determination of $\localDM$ from the rotation curve of our Galaxy, we analyze a selection of models with different mass distributions.
All the mass models that we study contain two main components: the baryonic part 
(in the form of disks and a bulge) 
and a dark matter halo. 
Apart from axisymmetry, we also assume in our analysis that the Galaxy is in a state of dynamical equilibrium.  Below we describe the baryonic and dark matter halo components that we investigate in our study and the motivation behind these choices. We first lay out all the considered baryonic models, followed by those of the dark matter halo.

\subsection{Baryonic models}\label{subsec:baryonic-models}
One of the fundamental tasks of Galactic Astronomy is to determine the luminosity distribution of the different baryonic components of the Milky Way---the stars, the gas and the dust. Their luminosity profiles effectively deliver their mass profiles. Various studies have been dedicated to this endeavour, but we yet lack a consensus in the baryonic model for the Milky Way 
(see e.g. \cite{Freeman:2002wq,Bland-Hawthorn:1602.07702}). 
For this reason, we iterate our analysis over two different baryonic models that are backed up by different observational studies. The choice of two different models also allows us to examine the effect that the baryonic matter distribution has on the estimates of $\localDM$.

\subsubsection{Baryonic model B1}\label{subsubsec:model-1}
We refer to our first mass model of baryons as B1. It essentially corresponds to the baryonic part of the Galactic model used by \cite{Eilers:1810.09466}, that they fix in their analysis. However, we consider both the case of parameters fixed to agree with their work and also the more general case in which we allow parameters to vary.
This model is based on Model I of \cite{Pouliasis:1611.07979}. It consists of three components, namely, a bulge and two disks (divided in terms of a thin and a thick distribution). The bulge is modeled as a Plummer potential \citep{Plummer:1911zza} given as
\begin{equation}\label{eq:Plummer}
\Phi_{\rm Plummer} (r) = - \frac{G M_{\rm bulge}}{\sqrt{r^2 + r_b^2}},
\end{equation}
where $r$ is the distance in galactocentric Cartesian coordinates and $r_b$ is the cut-off radius. The thin and the thick disks are represented independently by Miyamoto-Nagai potentials \citep{Miyamoto:1975zz} that are expressed as
\begin{equation}
\Phi_{\rm MN}(R, z) = - \dfrac{GM_{\rm disk}}{\sqrt{R^2+\left(R_d+\sqrt{z^2+z_d^2}\right)^2}} ,
\end{equation}
where $R$ is the radius in the galactocentric cylindrical coordinates, and $R_d$, $z_d$ are the characteristic scales of the profile. 

Whenever we fix the baryonic components of this model in our analyses, we set all the parameters at the same values used by \cite{Eilers:1810.09466}, as described in \cite{Pouliasis:1611.07979}, with $M_{\rm bulge} = 1.067\times 10^{10}\msun$ and $r_b=0.3\kpc$ for the bulge, and $M_{\rm thin} = M_{\rm thick} = 3.944\times 10^{10}\msun$, $R^{\rm thin}_d = 5.3 \kpc$, $z^{\rm thin}_d = 0.25\kpc$, $R^{\rm thick}_d = 2.6\kpc$ and $z^{\rm thick}_d = 0.8\kpc$ for the disks. 
The same parametric setting is used for the mean values of the Gaussian priors when we allow the baryonic components to vary in the fits.
The chosen values of the baryonic components are motivated by several observations, among which \cite{Pouliasis:1611.07979} used the local baryonic surface and volume densities of the Milky Way. The used observations are based on different studies \citep{Bensby:2011ae, Bovy:2011ux, Bland-Hawthorn:1602.07702, Kafle:2014xfa, McKee:2015hwa, Flynn:2006tm}. As we show in section~\ref{sec:analyses}, we study both the cases where we fix these parameters and also those in which we allow some variation. 

\subsubsection{Baryonic model B2}\label{subsubsec:model-2}

Miyamoto-Nagai disks have density profiles that do not decline as rapidly with radius as observed disk galaxies, and are expected to overestimate the mass of the baryons towards the outer Galactic radii \citep{Rix:2013bi}. For this reason, we study another baryonic model that we refer to as B2. 

The motivation for B2 comes mainly from the observational studies of \cite{Misiriotis:2006qq}, where they analyzed COBE dust emission maps \citep{Sodroski:1993mn,Fixsen:1996nj} to constrain the parameters of their Galactic model. Such model is comprised of axisymmetric distributions for the stars (composed of a bulge and a disk), the dust (cold and warm) and the gas (molecular $\mathrm{H}_2$ and atomic HI). The stellar disk, the two dust components and the $\mathrm{H}_2$ gas distribution are all modelled as double exponential profiles expressed as
\begin{equation}\label{eq:doubleexponential}
    \rho (R,z) = \rho_0 \exp \left( -\frac{R}{R_d} - \frac{|z|}{z_d} \right),
\end{equation}
where $\rho_0 = M/(4\pi z_d R_d^2)$ is the normalization, $M$ is the corresponding mass, and $R_d$ and $z_d$ are the characteristic scale length and height, respectively. 
We set these parameters as $M_{\rm disk} = 3.65 \times 10^{10}\,\mathrm{M}_\odot$ \citep{Portail:1502.00633,Bland-Hawthorn:1602.07702}, $R_d^{\rm disk} = 2.35\kpc$, $z_d^{\rm disk} = 0.14\kpc$; $M_{\rm cold} = 7.0\times 10^7 \msun$, $R_d^{\rm cold} = 5.0\kpc$, $z_d^{\rm cold} = 0.1\kpc$; $M_{\rm warm} = 2.2\times 10^5 \msun$, $R_d^{\rm warm} = 3.3\kpc$, $z_d^{\rm warm} = 0.09\kpc$, and $M_{\rm H_2} = 1.3\times 10^9 \msun$, $R_d^{\rm H_2} = 2.57\kpc$, $z_d^{\rm H_2} = 0.08\kpc$ \citep{Misiriotis:2006qq}.
The atomic HI gas distribution is also modelled as a double exponential, as per Eq.~\eqref{eq:doubleexponential}, with its mass defined as $M = 4\pi \rho_0 z_d R_d (R_t + R_d)\, \mathrm{e}^{-R_t / R_d}$, where $M_{\rm HI} = 8.2\times 10^9 \msun$, $R_d^{\rm HI} = 18.24\kpc$, $z_d^{\rm HI} = 0.52\kpc$ and $R_t = 2.75\kpc$, as obtained in \cite{Misiriotis:2006qq}. This mass definition corresponds to that of a truncated disk, although we are using a double exponential shape without truncation for the sake of easier numerical analyses. This is feasible because the contribution that gas and dust terms adds to the global circular velocity curve is significantly smaller than that of the remaining components. Therefore, this minor approximation does not cause any marked effect. We however include these components in our total baryonic mass model for consistency and in order to have a more realistic distribution of baryons.

For the bulge, we substitute the original profile used in \cite{Misiriotis:2006qq} with a Hernquist potential that is given as
\begin{equation}\label{eq:Hernquist}
    \Phi_{\rm Hernquist} (r) = - \frac{G M_{\rm bulge}}{r_b + r}, 
\end{equation}
where we set $M_{\rm bulge} = 1.55 \times 10^{10}\,\mathrm{M}_\odot$ \citep{Portail:1502.00633,Bland-Hawthorn:1602.07702}\footnote{The value of $M_{\rm bulge}$ estimated by \cite{Portail:1502.00633} was obtained from a triaxial modelling of the bulge region.} and $r_b = 0.7\kpc$. 
This change is also motivated by an easier numerical calculation, since the original profile used in \cite{Misiriotis:2006qq} has a more complicated shape. We decided to use a Hernquist profile (Eq.~\eqref{eq:Hernquist}) instead of a Plummer potential (Eq.~\eqref{eq:Plummer}) because a Hernquist profile with $r_b = 0.7\kpc$ adjusts better, given the same mass, to the bulge's shape used in \cite{Misiriotis:2006qq}.
In our analysis, the values of $M_{\rm bulge}$ and $M_{\rm disk}$ (or equivalently the central values of their Gaussian prior distributions, when these parameters are allowed to vary) were chosen such that $M_{\rm bulge}/(M_{\rm bulge}+M_{\rm disk} ) = 0.3$ \citep{Bland-Hawthorn:1602.07702}.

\subsection{Dark matter profiles}\label{subsec:DM-models}
One of the long-standing problems of astrophysics and cosmology has been to determine the dark matter distribution around the Milky Way. 
Since a change of the dark matter halo can affect the estimate of $\localDM$ from the Milky Way's rotation curve, we use two different parameterizations for the dark matter component, allowing for different shapes that deviate particularly towards the centre. Specifically, one of the profiles that we use presents a cusp in the centre (see Eq.~\eqref{eq:rho-MNFW} below) while the other one can exhibit a cored halo (see Eq.~\eqref{eq:rho-Ein} below) for some values of its free parameter $\alpha$.

\begin{enumerate}
    \item We describe our first dark matter halo as a generalized version of the Navarro-Frenk-White (NFW) profile, that we refer to as gNFW,
\begin{equation}\label{eq:rho-MNFW}
    \rho_{\gNFW} (r) = \rho_0 \left( \frac{r_s}{r} \right)^\gamma \left( 1 + \frac{r}{r_s} \right)^{\gamma - 3},
\end{equation}
where $\rho_0$ is the normalization constant, $r_s$ the scale radius and $\gamma$ the inner slope. For $\gamma=1$ the profile becomes the well-known NFW halo 
($\rho_{\rm NFW}$, \cite{Navarro:1996gj}). 
This profile is a common approximation to dark matter densities found in cosmological simulations, and it represents a cuspy profile that diverges towards smaller $r$ values. We incorporate both cases in our study, one where $\gamma$ is a free parameter (varying in the range $\gamma = \text{0--2}$) and another where we set $\gamma=1$ (which then takes the form of the usual NFW).

\item The second dark halo that we consider is the Einasto profile \citep{Einasto:1965czb}, which is expressed as
\begin{equation}\label{eq:rho-Ein}
    \rho_{\rm Ein} (r) = \rho_0 \exp \left\{ -\frac{2}{\alpha} \left( \left( \frac{r}{r_s} \right)^\alpha - 1 \right) \right\},
\end{equation}
where $\alpha$ determines how fast the density distribution falls with $r$, making it cored towards the central regions of the Galaxy when its value is close to $\alpha = 1$. We allow for values of this parameter in the range $\alpha = \text{0--1}$.

\end{enumerate}

In the next section, the parameters of the dark matter halo that we will directly constrain in our analyses (comparing the Milky Way's circular velocity curve data from \cite{Eilers:1810.09466} with the outcome of our Galactic mass models) are the virial mass $M_{200}$, the virial concentration $c_{200}$, and the $\gamma$ and $\alpha$ parameters.
With this definition, $M_{200}$ is the mass contained within the radius $r_{200}$ such that the energy density is 200 times larger than the critical energy density $\rho_{\rm crit}$ of the Universe,
\begin{equation}\label{eq:M200-def}
M_{200} = \frac{4\pi}{3} r^3_{200} \Delta_{200} \rho_{\rm crit},
\end{equation}
where $\Delta_{200}= 200$.
Notice that this is just a matter of definition, so choosing this or another virial overdensity (like the one derived from the collapse of a spherical top-hat 
perturbation \cite{Bryan:1997dn}) 
does not affect our results. 

One remark worth making is  about the concentration parameter, defined as
\begin{equation}\label{eq:c200-def}
c_{200} = \frac{r_{200}}{r_{-2}} ,    
\end{equation}
where $r_{-2}$ is the radius at which the slope $\mathrm{d}\,\mathrm{ln}\rho / \mathrm{d}\,\mathrm{ln}r = -2$. In the  $\rho_{\gNFW}$ case, $r_{-2} = (2-\gamma ) r_s$. This means that for very steep $\gNFW$ halos, with values of $\gamma$ close to 2, $r_s$ becomes extremely large. On the other hand, for  $\rho_{\rm Ein}$ we have the relation $r_{-2} = r_s$. 

Defining $\eta$ as the parameter $\gamma$ for the $\gNFW$ profile and $\alpha$ for the Einasto profile, the connection between the varied dark matter quantities of our analyses, $\left\{M_{200},c_{200},\eta\right\}$, and a given dark matter profile $\rho_{\rm DM}(r,r_s,\rho_0,\eta)$, is made through Eqs.~\eqref{eq:M200-def}, \eqref{eq:c200-def} and
\begin{equation}\label{eq:M200-norm-integral}
M_{200} = 4\pi \int_0^{r_{200}} \rho_{\rm DM}(r',r_s,\rho_0,\eta)\, r'^2 \,\mathrm{d}r'.
\end{equation}
For each set of values $\left\{ M_{200}, c_{200}, \eta \right\}$ we first determine $r_{200}$ using Eq.~\eqref{eq:M200-def}, then we obtain $r_s$ from Eq.~\eqref{eq:c200-def}, and finally we get the value of the normalization $\rho_0$ that determines $\rho_{\rm DM}(r,r_s,\rho_0,\eta)$ solving Eq.~\eqref{eq:M200-norm-integral}.

\section{Analyses}\label{sec:analyses}

In this section, we present our analyses of the Milky Way's rotation curve based on the implementation of the aforementioned Galactic mass models. To this end, we perform Bayesian analyses, comparing the circular velocity measurements against the circular velocity values obtained from the mass models, in order to constrain $\localDM$. In particular, we run a Markov Chain Monte Carlo (MCMC) using the affine invariant sampler \texttt{emcee} \citep{ForemanMackey:2012ig} implemented in Python, with the log-likelihood function defined by
\begin{equation}
     \ln \mathcal{L} = -\frac{1}{2}\sum_{i} \left( \frac{v_{\rm c}^{\rm d} (R_i) - v_{\rm c}^{\rm m}(R_i) }{\sigma_i} \right)^2 ,
\end{equation}
where the summation $i$ is done over all the data points.
$v_{\rm c}^{\rm d}(R_i)$ is the measurement of the circular velocity at a given radius $R_i$ in the galactocentric cylindrical coordinates, and the corresponding model value is given by $v_{\rm c}^{\rm m}(R_i)$. 
The Gaussian dispersion $\sigma_i$ is quadratically summed over the statistical and systematic uncertainties associated to the $i$-th data point, 
where the statistical uncertainties are taken from Tab. 1 of \cite{Eilers:1810.09466} and the systematic uncertainties are extracted from their Fig. 4, assuming for the last bins, which are saturated in the figure, a 12\% deviation that corresponds to the maximum deviation shown in that plot. 
When quadratically summing both statistical and systematic uncertainties we are 
simplifying the treatment of systematic uncertainties. 
In particular, we are adding extra freedom that is not necessarily allowed by the data, since in principle there are sources of systematic uncertainties that imply some correlation in the radial direction. 
Alternatively, a better approach could be to use a Gaussian process, or to repeat the analyses by systematically shifting the data points within the allowed range of systematic uncertainties, with $\sigma_i$ being only the statistical uncertainties in the second case.
Although the estimated systematic error that we get might be larger than the real error caused by systematic uncertainties, our chosen treatment is simple from a computational point of view and it provides a reasonable result, since the size of systematic uncertainties is not large enough to produce a dangerous discrepancy between the different approaches.

Given a theoretical mass model, we evaluate $v_{\rm c}^{\rm m}(R)$ as
\begin{equation}
    v_{\rm c}^{\rm m}(R) = \left( v_{\text{c,DM}}^{2}(R) + v^2_{\rm c,B}(R) \right)^{1/2} ,
\end{equation}
where $v^2_{\text{c,DM}}(R)$ is the contribution to the circular velocity coming from dark matter and $v^2_{\rm c,B}(R)$ is the summed contribution of the different baryonic components, as described in section~\ref{sec:models}.
The circular velocity for each component is obtained from its potential using Eq.~\eqref{eq:vc-from-potential}.

Furthermore, the uncertainty in the estimate of $\localDM$ is related to the dependence of this quantity upon the solar distance from the Galactic centre, $R_\odot$. Since the value of $R_\odot$ also has some associated uncertainty, it is something that needs to be taken into account. Therefore, $R_\odot$ is included as an additional parameter over which we sample in our MCMC. In particular, we use $R_\odot = 8.122\pm 0.031\,\mathrm{kpc}$ \citep{GravityColl:2018}, which is the same value adopted by \cite{Eilers:1810.09466}. Larger uncertainties are usually adopted for this quantity 
(see e.g. \cite{Bland-Hawthorn:1602.07702}). 
However, throughout our study we observe that the estimates of $\localDM$ are mostly susceptible to changes in the underlying distribution of baryons.


In all our analyses we consider flat priors for the dark halo parameters: $M_{200}$, $c_{200}$, $\alpha$ and $\gamma$, that can vary in the ranges 
$M_{200}/[10^{11}\msun]\in [0.01,30]$, $c_{200} \in [0,100]$, $\alpha \in [0,1]$ and $\gamma \in [0,2]$. We also tested that choosing a broader flat prior in $\log_{10} M_{200}$ does not affect our results.

\subsection{Analysis of model B1}\label{subsec:analysis-B1}
We first study our baryonic model B1, defined in section \ref{subsubsec:model-1}, in combination with different dark halo profiles, to find estimates for $\localDM$ from matching to the circular velocity curve data.  We begin in section \ref{subsubsec:Eilers-baryons-fixed} by studying exactly the same case as done in \cite{Eilers:1810.09466}, using their values of baryonic parameters together with a standard NFW halo. Our resulting $\localDM$ agrees with theirs, but we find even smaller error bars for this particular case, as we will explain.  Then in section \ref{subsubsec:beyond-Eilers-baryons-masses} we generalize by allowing the values of the baryonic parameters to vary, 
and in section \ref{subsubsec:changing-DM-profile} we generalize further by considering a variety of dark matter halo profiles.

\subsubsection{Fixing baryonic parameters as in the case studied by \protect\cite{Eilers:1810.09466}}
\label{subsubsec:Eilers-baryons-fixed}

In this section we first study the value of $\localDM$ obtained with exactly the same mass model as in the illustration case presented by \cite{Eilers:1810.09466}. We use the model B1 for the baryonic components with the parameters fixed to the values presented in section \ref{subsubsec:model-1}, and we use an NFW profile for the dark halo ($\rho_{\gNFW}$ with inner slope $\gamma = 1$), thereby fitting for the NFW parameters $M_{200}$ and $c_{200}$. In addition, we distinguish two cases, one in which we do not include the systematic uncertainties of the data in the fit and another in which we include them.
We make this distinction because we are able to reproduce the results of \cite{Eilers:1810.09466} only when systematics are not included. 

\begin{figure*}
\begin{center}
\includegraphics[height=0.22\textheight]{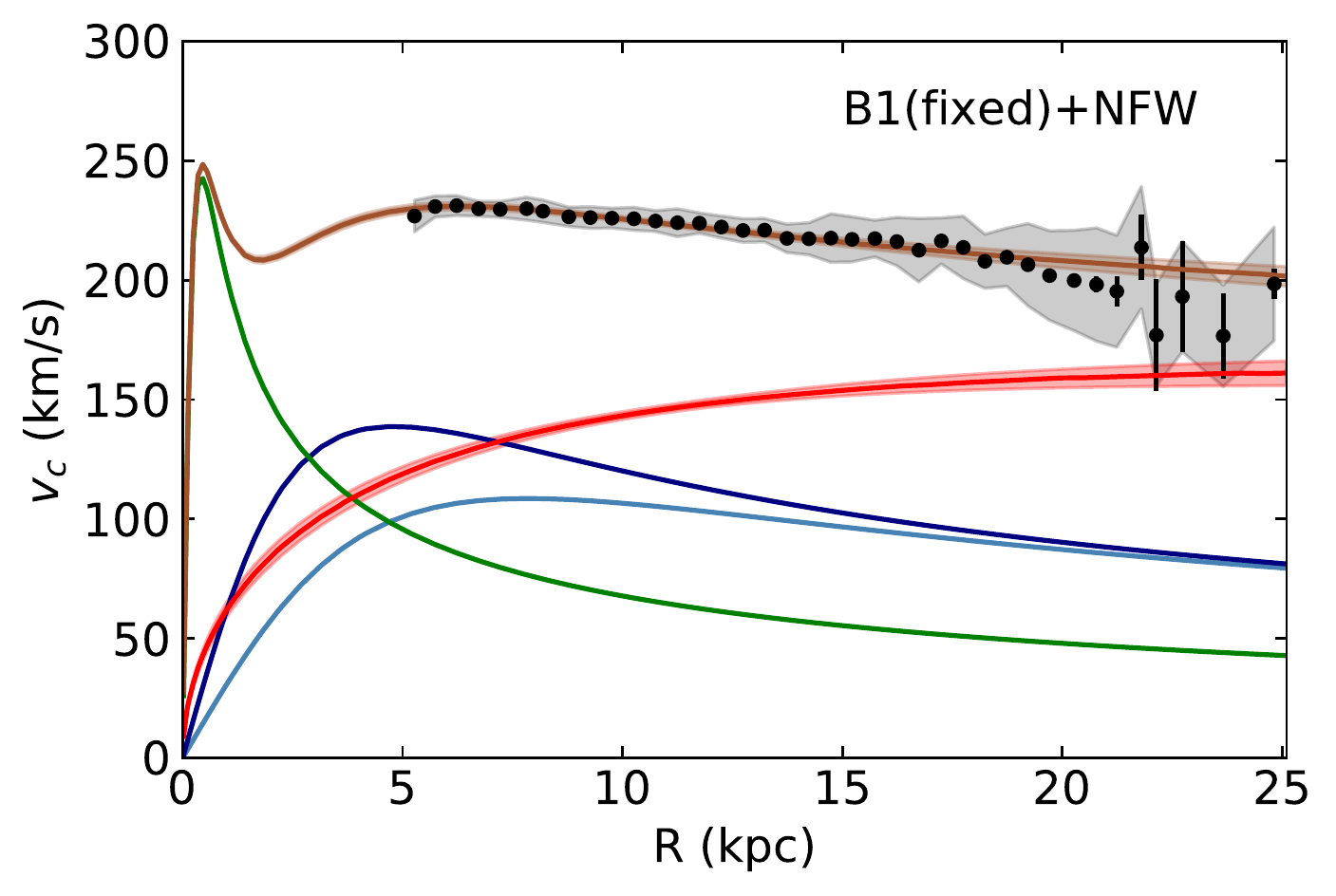}
\includegraphics[height=0.22\textheight]{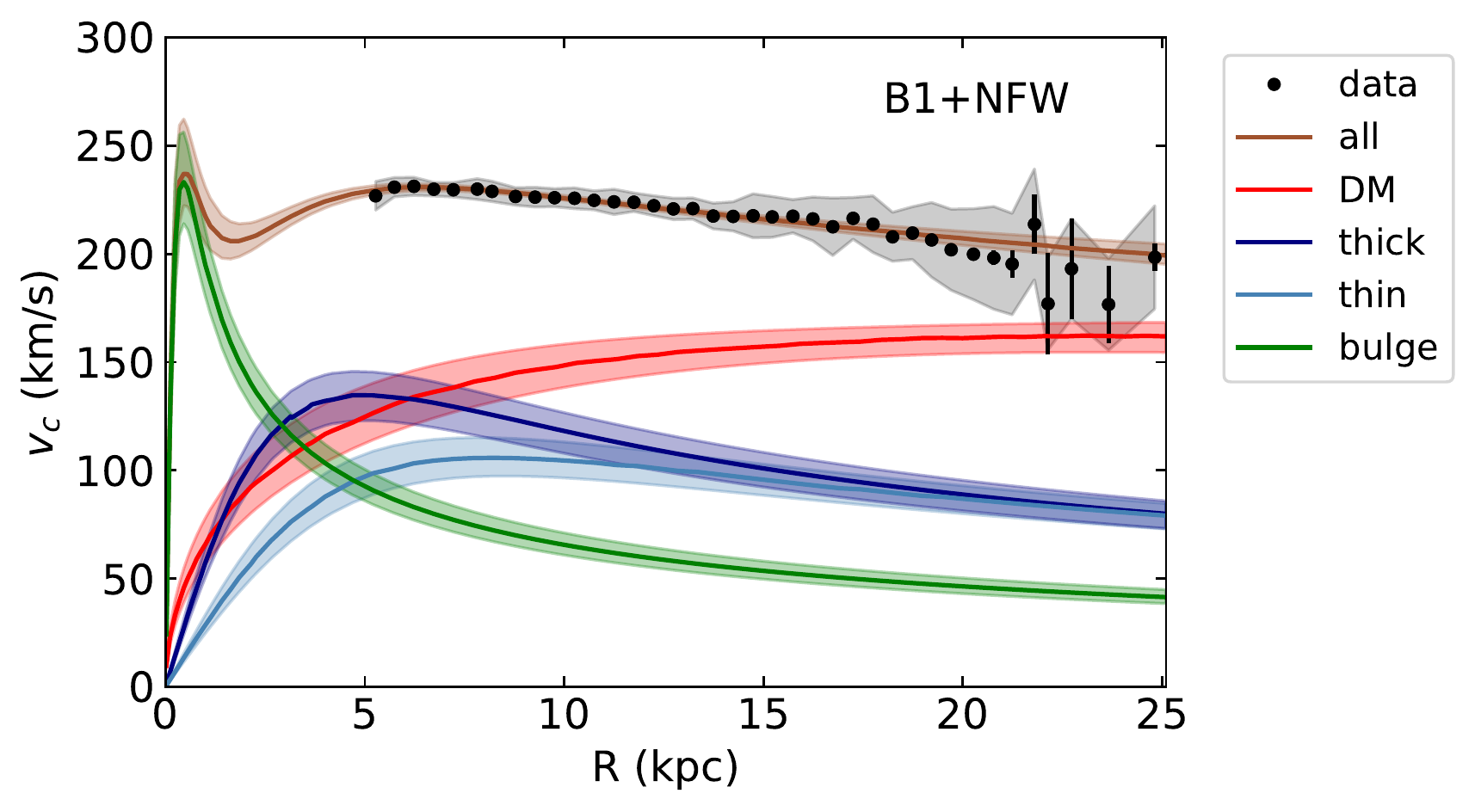}
\end{center}
\caption{Circular velocity curves for the case of an NFW dark halo and baryons from the model B1.  \textit{Left panel}: Parameters of the baryonic components fixed to the values described in section \ref{subsubsec:model-1} (the same mass model as \protect\cite{Eilers:1810.09466}). \textit{Right panel}: Baryonic components allowed to vary with Gaussian priors as described in section~\ref{subsubsec:beyond-Eilers-baryons-masses}. The black shaded region corresponds to $1\sigma$ systematic uncertainties and other areas correspond to marginal 68\% credible regions, with the mean values shown as lines.}
\label{fig:Eilers-rep}
\end{figure*}

\begin{figure}
    \centering
    \includegraphics[width=0.45\textwidth]{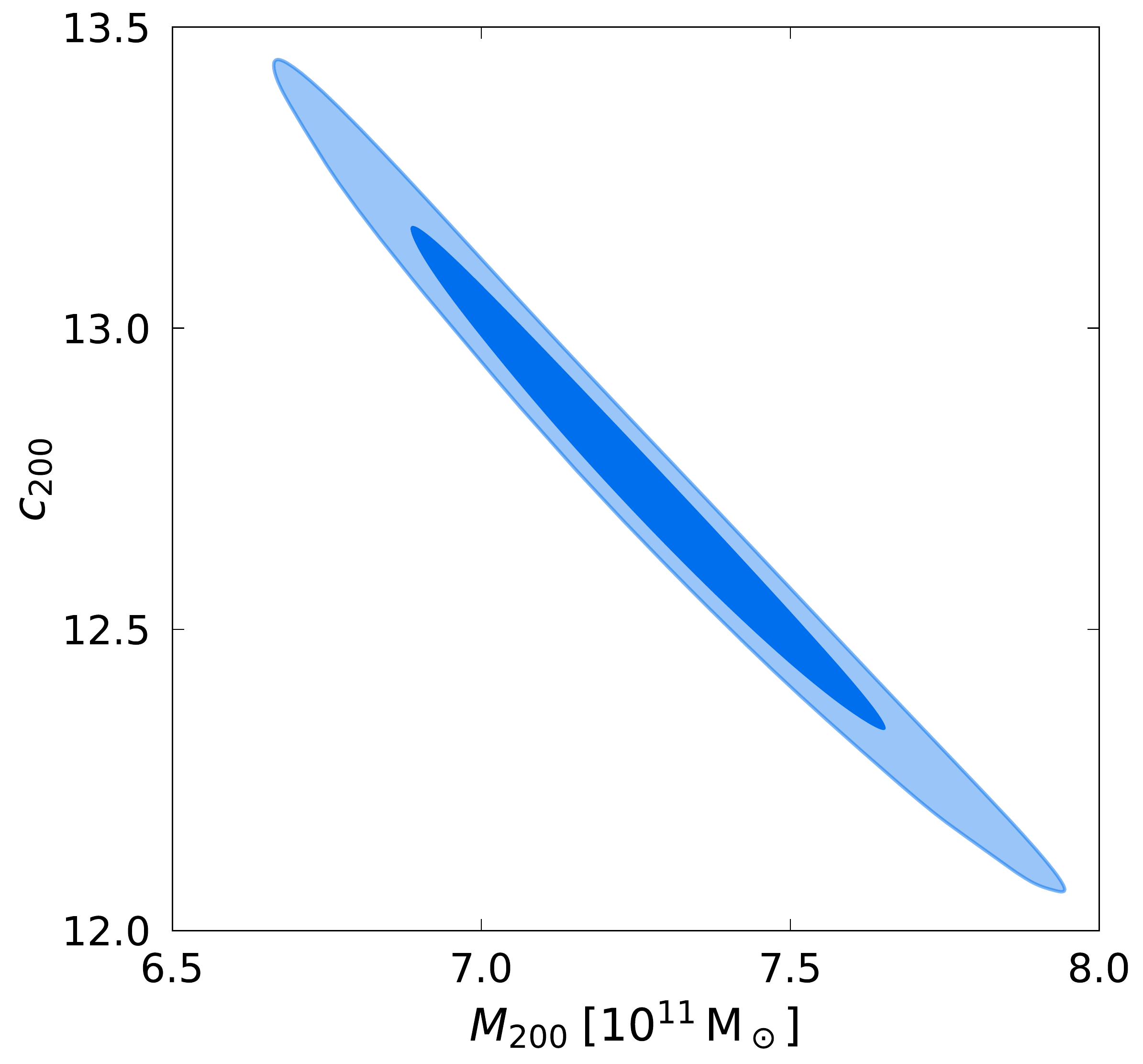}
    \caption{Correlation found between the two fitted virial quantities when we analyze the baryonic model B1 and an NFW dark matter halo (with the same parameter choices as were studied by \protect\cite{Eilers:1810.09466}), with no systematic uncertainties in the data and when baryonic parameters are fixed to the values given in \protect\cite{Pouliasis:1611.07979}. 
    The 68\% and 95\% credible regions are shown.
    This correlation is independent of the $\cmrelation$ relation found in $N$-body simulations, that we discuss in section~\ref{subsec:fit-cvir-Mvir-relation}.
    }
    \label{fig:MvirCvir-rep}
\end{figure}

The best-fit circular velocity curve when we include systematic uncertainties is shown in the left panel of Fig.~\ref{fig:Eilers-rep}, together with the individual contribution from each Galactic component. The corresponding fitted values for the parameters of the dark matter profile are provided in Tab.~\ref{tab:Eilers-fixed}. Comparing these results with those of \cite{Eilers:1810.09466}, we find that we are successfully able to reproduce their analysis only when no systematics are included, with the single exception of the uncertainty associated with the derived $\localDM$, for which we find smaller error bars. 
We think that the reason is due to a correlation found in our analysis between the two fitted quantities $M_{200}$ and $c_{200}$ (see Fig.~\ref{fig:MvirCvir-rep}). 
Since $\localDM$ is a derived parameter that depends on $M_{200}$, $c_{200}$ and $R_\odot$, one needs to propagate the error of the virial quantities in order to infer the error of $\localDM$. If we neglect the correlation shown in Fig.~\ref{fig:MvirCvir-rep} and compute the uncertainty of $\localDM$ from quadratic error propagation of $R_\odot$, $M_{200}$ and $c_{200}$, assuming that they are independent, we get an uncertainty that is consistent with \cite{Eilers:1810.09466} findings.
It is important to highlight that we correct this aspect and account for the correlation in the rest of the study.

\begin{table}
    \centering
    \begin{tabular}{l|c|c}
    \hline
 & without systematics & with systematics \\
    \hline
$M_{200}\;[10^{11}\,\mathrm{M}_\odot]$ & $7.25\pm 0.26$ & $6.5_{-1.2}^{+1.6}$ \\[\tabSpace]
$c_{200}$ & $12.7\pm 0.3$ & $12.9_{-1.5}^{+1.7}$ \\
    \hline
$\localDM \; [\mathrm{GeV/cm^3}]$ & $0.2981_{-0.0023}^{+0.0022}$ & $0.293_{-0.010}^{+0.009}$ \\[\tabSpace]
$r_{200}\;[\mathrm{kpc}]$ & $189.4_{-2.3}^{+2.1}$ & $184_{-12}^{+14}$ \\[\tabSpace]
$r_{s}\;[\mathrm{kpc}]$ & $14.9\pm 0.5$ & $13.7_{-2.3}^{+3.1}$ \\
    \hline
    \end{tabular}
    \caption{Fitted and derived quantities obtained from the analysis of the baryonic model B1 when their parameters are fixed, as in the analysis of $\localDM$ presented by \protect\cite{Eilers:1810.09466}, and an NFW dark matter halo (section~\ref{subsubsec:Eilers-baryons-fixed}). Top rows correspond to parameters of the fit, while the last rows are derived quantities. The values correspond to the maximum and 68\% credible region of the marginal posteriors.}
    \label{tab:Eilers-fixed}
\end{table}

\subsubsection{Varying baryonic parameters}
\label{subsubsec:beyond-Eilers-baryons-masses}

In the previous section we fixed the baryonic components of the Galaxy, but our knowledge on the distribution of baryons in the Milky Way is not perfect. Therefore, as a first approach to test the dependence of $\localDM$ on the baryonic parameters, we free our model B1 to let it vary in our analysis. However, in order to retain the essence of the model, we impose a Gaussian prior on the parameters, taking as central values those discussed in section~\ref{subsubsec:model-1}, with a standard deviation corresponding to 15\% of the central values for the masses and 10\% for the characteristic scales. These standard deviations enter in the range of uncertainties presented in \cite{Bland-Hawthorn:1602.07702,Pouliasis:1611.07979}. As in the previous section, we assume a standard NFW halo for the dark matter.

The circular velocity curves obtained from this analysis are shown in the right panel of Fig.~\ref{fig:Eilers-rep} and the corresponding dark matter halo fitted parameters are listed in the first column of Tab.~\ref{tab:Eilers-varying-othersDM}. Comparing these results with the values of Tab.~\ref{tab:Eilers-fixed} we can see that the error bars of all the dark matter parameters broaden when baryons are allowed to change.

In addition, we noticed that the resulting posterior distributions of the baryonic parameters are dominated by their priors. This is an expected behaviour for two reasons: \emph{a)} the constraining power of $v_{\rm c}$ is not enough to fit all the baryonic parameters together with those of the dark matter, and \emph{b)} the baryonic contribution to $v_{\rm c}$, for most of the range of $R$ values covered by the data, is smaller than the contribution of the dark matter component, as can be realized from the circular velocity curve figures. We also checked that doubling the uncertainties on the baryonic parameters does not make a significant difference in the $\localDM$ estimate, and neither to its uncertainty. However, when we imposed flat priors on our baryonic model, we found that the analysis preferred to fit the data with the dark matter distribution alone, ending up making the contribution from baryons too small to be realistic. This already points towards the dependence of the estimated $\localDM$ on the assumed baryonic distribution.

\subsubsection{Changing the dark matter profile}
\label{subsubsec:changing-DM-profile}

We now study the effect that changing the dark matter profile has on the inferred value of $\localDM$.  
We use the same varying baryonic parameters for model B1 as in section~\ref{subsubsec:beyond-Eilers-baryons-masses}; however, for the dark matter profile, we now generalize from standard NFW to $\gNFW$ with $\gamma$ being a free parameter. In other words we now let the inner slope $\gamma$ of the $\rho_{\gNFW}$ dark matter halo  (Eq.~\eqref{eq:rho-MNFW}) become a free parameter of the mass model. The values of the dark matter fitted and derived quantities are shown in Tab.~\ref{tab:Eilers-varying-othersDM}.   One can see that the resulting value of $\localDM$ is essentially the same as before, despite the freedom in the dark matter profile.
However, the additional inclusion of another free parameter in general broadens the obtained uncertainties of the velocity curves. The standard NFW profile with $\gamma =1$ is well contained within the 68\% credible region when $\gamma$ is allowed to change. 
 
On the other hand, the values of the baryonic parameters are not affected by this change in the dark matter profile, since they are still prior dominated (see the discussion in section~\ref{subsubsec:beyond-Eilers-baryons-masses}). 

On a similar basis, running the analysis by implementing an Einasto profile to model the dark matter distribution (Eq.~\eqref{eq:rho-Ein}) also has a negligible effect on the estimate of $\localDM$ (see Tab.~\ref{tab:Eilers-varying-othersDM}). Since this profile is naturally less peaked at the centre, the virial mass is reduced by about a factor of two with respect to the standard and generalized NFW studies. However, their values are compatible within their uncertainties. 

In Fig.~\ref{fig:triang-rel-B1-NFW} of the appendix~\ref{appendix:triang-plots} we show the plot including the 2D marginal credible intervals for the parameters of Tab.~\ref{tab:Eilers-varying-othersDM}.

\begin{table}
    \centering
    \begin{tabular}{l|c|c|c}
    \hline
 & NFW & $\gNFW$ & Einasto \\
    \hline
$M_{200}\;[10^{11}\,\mathrm{M}_\odot]$ & $5.2^{+2.0}_{-1.1}$ & { $5.5_{-1.4}^{+3.1}$} & $2.8_{-1.2}^{+7.7}$ \\[\tabSpace]
$c_{200}$ & $15^{+5}_{-4}$ & { $14\pm 5$} & $12\pm 4$ \\[\tabSpace]
Slope parameter & $\gamma = 1$ & $\gamma = 1.2_{-0.8}^{+0.3}$ & $\alpha = 0.11_{-0.05}^{+0.20}$ \\
    \hline
$\localDM \; [\mathrm{GeV/cm^3}]$ & $0.301^{+0.028}_{-0.025}$ & { $0.300_{-0.027}^{+0.028}$} & $0.301\pm 0.027$ \\[\tabSpace]
$r_{200}\;[\mathrm{kpc}]$ & $173^{+19}_{-13}$ & { $174_{-15}^{+29}$} & $182_{-51}^{+43}$ \\[\tabSpace] 
$r_{s}\;[\mathrm{kpc}]$ & $10^{+5}_{-3}$ & { $9_{-8}^{+12}$} & $11_{-4}^{+10}$ \\
    \hline
    \end{tabular}
    \caption{Dark matter related quantities obtained from the fit of the baryonic model B1, when the parameters of the baryonic components are allowed to vary, using an NFW, a $\gNFW$ and an Einasto dark matter halo. For the last two columns $\gamma$ and $\alpha$ are extra free parameters of the analysis. Top rows correspond to fitted variables and the last rows are derived quantities. The values correspond to the maximum and 68\% credible region of the marginal posteriors.}
\label{tab:Eilers-varying-othersDM}
\end{table}
\subsection{Analysis of model B2}\label{subsec:analysis-B2}

We now repeat the study presented in the previous section, although this time modelling the baryonic components with those of the model B2 (see section~\ref{subsubsec:model-2} for the details). The fact that the resulting $\localDM$ is driven by our choice of the baryonic model will become clearer in this section.

We again allow baryons to change with Gaussian priors as described before (see section~\ref{subsubsec:beyond-Eilers-baryons-masses}), with the exception of the parameters for the gas distribution, for which we consider standard deviations corresponding to 25\% and 20\% from the central values of their masses and characteristic scales, respectively. The reason for this broadening is that these components are distributed in a clumpier way in the Galaxy, so the axisymmetric profiles provide a worse fit to the COBE maps in the analysis of \cite{Misiriotis:2006qq}.

As we did in our study of the baryonic model B1, we start by assuming an NFW dark halo in section~\ref{subsubsec:fit-B2-NFW} and change it later in section~\ref{subsubsec:fit-B2-other-DM}.

\subsubsection{NFW dark matter profile}\label{subsubsec:fit-B2-NFW}

\begin{figure}
    \centering
    \includegraphics[width=0.49\textwidth]{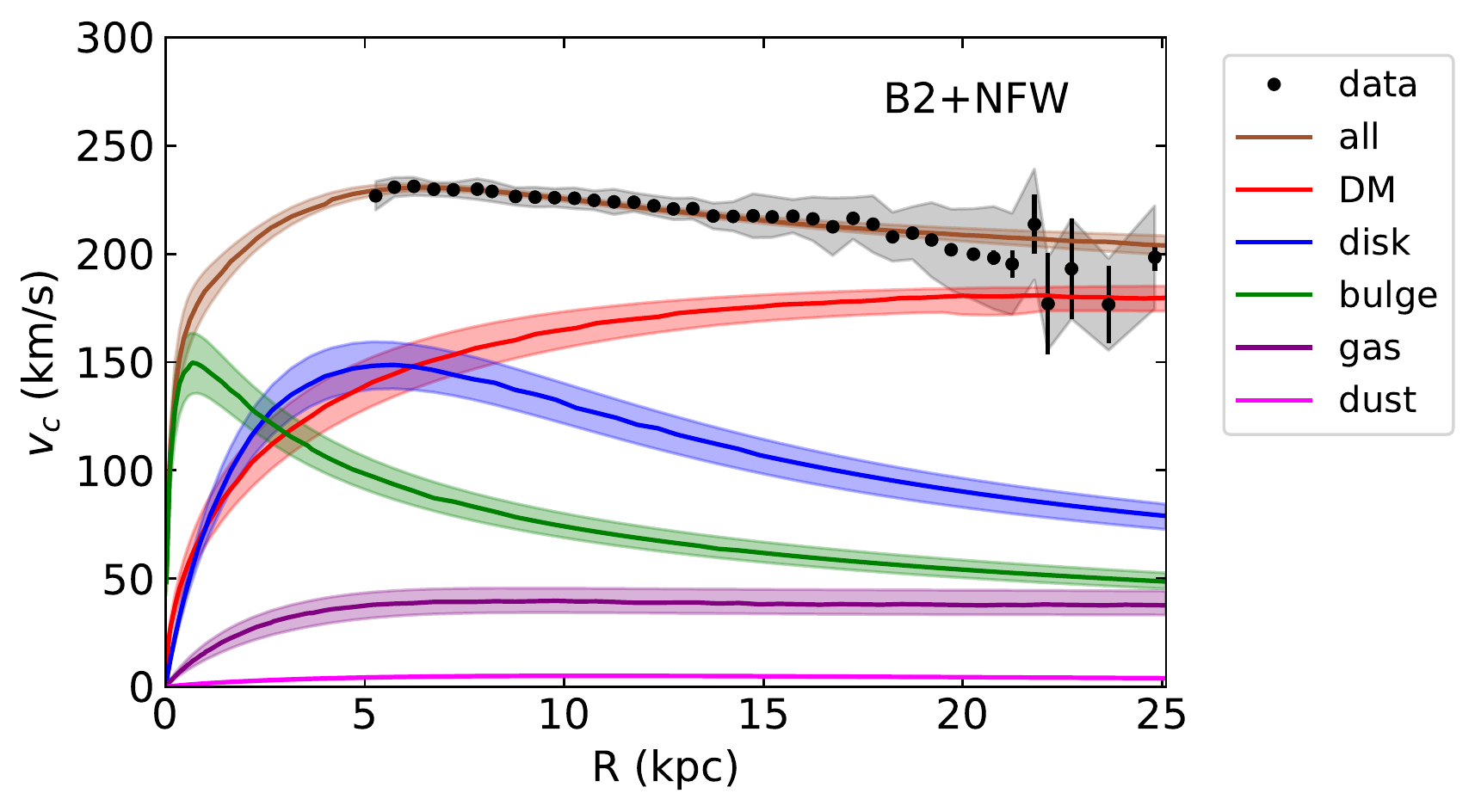}
    \caption{Circular velocity curves for the analysis of our baryonic model B2 and an NFW dark matter profile (see section~\ref{subsubsec:fit-B2-NFW}). Similar results are obtained when the dark matter halo is changed to a $\gNFW$ or an Einasto (see Fig.~\ref{fig:B2-MNFW-Einasto}). The black shaded region corresponds to $1\sigma$ systematic uncertainties and other areas correspond to marginal 68\% credible regions, with the mean values shown as lines.}
    \label{fig:B2-NFW}
\end{figure}

The first dark halo that we study is the standard NFW halo ($\gNFW$ with $\gamma = 1$).  The resulting values of the dark matter parameters are presented in Tab.~\ref{tab:B2-NFW-MNFW-Einasto} (first column of results) and the associated circular velocity curves are shown in Fig.~\ref{fig:B2-NFW}.

Partly because the total baryonic mass of the B1 model is larger than the baryonic mass of B2, the contribution to the circular velocity curve of baryons from the B2 model is smaller than the contribution of baryons from the B1 model. Therefore, a slightly more massive and concentrated dark halo is needed to fit the data well (compare the values of Tabs.~\ref{tab:Eilers-varying-othersDM} and \ref{tab:B2-NFW-MNFW-Einasto}). As a result, the estimated value of $\localDM$ in the analysis of the baryonic model B2 is about 30\% larger than in the analysis of the B1 model.

In principle, even though the mass of the stellar disk of model B2 is smaller than the mass of each of the two disks of model B1, the contribution to the circular velocity curve coming from the disk of model B2 is larger than the individual contributions of the baryonic disks of model B1, as can be appreciated comparing Figs.~\ref{fig:Eilers-rep} and \ref{fig:B2-NFW}. The difference is due to the larger density in the Galactic plane (at a height $z=0$) of the double exponential disk from the model B2, compared to the same density from the Miyamoto-Nagai disks of the B1 model.

The different choice of the bulge profile in the two baryonic models affects mostly the circular velocity at distances smaller than $5\kpc$ from the Galactic centre, having little effect on the estimated value of $\localDM$. Furthermore, observationally we can not associate properly a value of the circular velocity to measurements in the region $R\lesssim 5\kpc$ \citep{Chemin:1504.01507}. In addition, it is known that the bulge is not spherical at such distances 
(see e.g. \cite{Portail:1502.00633}), 
so a spherical approximation can be used as long as we do not include data at $R\lesssim 5\kpc$.

\begin{table}
    \centering
    \begin{tabular}{l|c|c|c}
    \hline
 & NFW & $\gNFW$ & Einasto \\
    \hline
$M_{\rm 200}\;[10^{11}\,\mathrm{M}_\odot]$ & $7.4_{-1.5}^{+1.8}$ & { $6.3_{-1.3}^{+3.4}$} & $3.0_{-1.2}^{+5.7}$ \\[\tabSpace]
$c_{\rm 200}$ & $16_{-3}^{+4}$ & { $17\pm 6$} & $14_{-4}^{+5}$ \\[\tabSpace]
Slope parameter & $\gamma = 1$ & $\gamma = 1.3_{-0.9}^{+0.3}$ & $\alpha = 0.18_{-0.09}^{+0.21}$ \\
    \hline
$\localDM \; [\mathrm{GeV/cm^3}]$ & $0.376\pm 0.025$ & { $0.387_{-0.036}^{+0.034}$} & $0.384_{-0.034}^{+0.038}$ \\[\tabSpace]
$r_{200}\;[\mathrm{kpc}]$ & $192_{-13}^{+15}$ & { $184_{-14}^{+29}$} & $147_{-19}^{+59}$ \\[\tabSpace]
$r_{s}\;[\mathrm{kpc}]$ & $11_{-3}^{+4}$ & { $8.1_{-7.8}^{+10.6}$} & $9.2_{-2.7}^{+5.3}$ \\
    \hline
    \end{tabular}
    \caption{Dark matter related quantities obtained from the fit of the baryonic model B2 with an NFW, a $\gNFW$ and an Einasto dark matter halo. Top rows correspond to fitted variables and the last rows are derived quantities. As usual, for the NFW halo the inner slope is fixed to be $\gamma = 1$. The values correspond to the maximum and 68\% credible region of the marginal posteriors.}
\label{tab:B2-NFW-MNFW-Einasto}
\end{table}

\subsubsection{Changing the dark matter profile}\label{subsubsec:fit-B2-other-DM}

\begin{figure*}
    \centering
    \includegraphics[height=0.22\textheight]{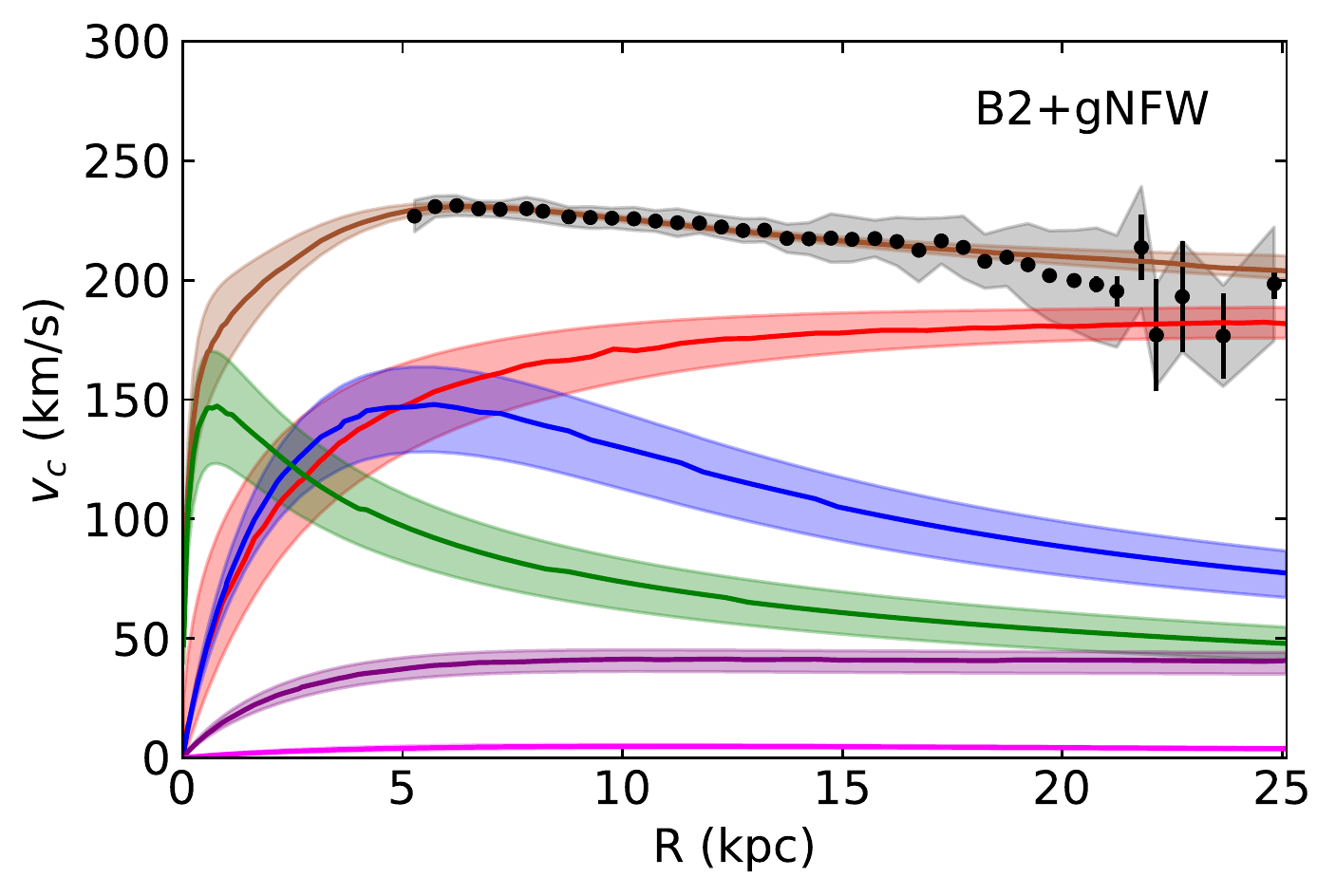}
    \includegraphics[height=0.22\textheight]{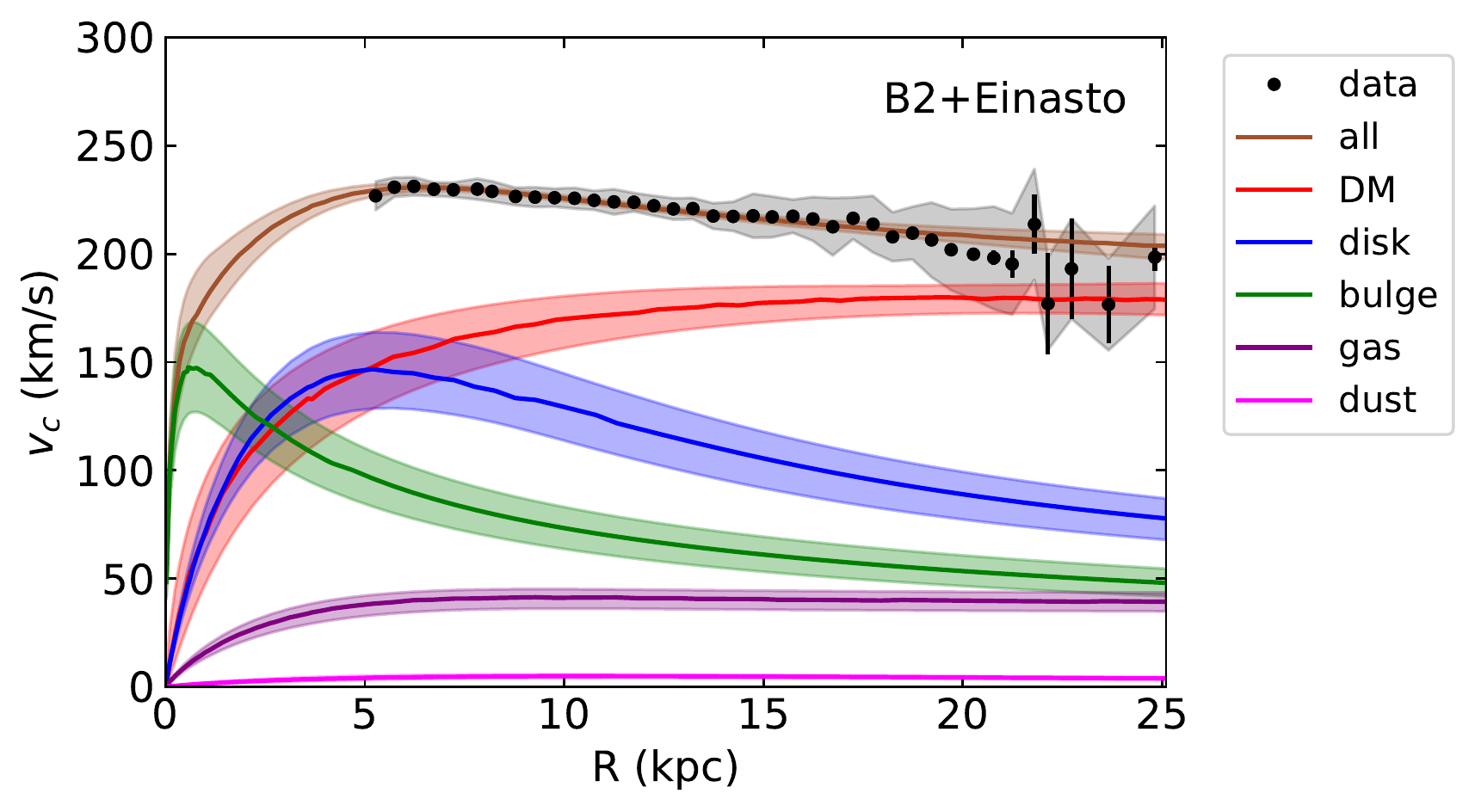}
	\caption{Circular velocity curves for the analysis of our baryonic model B2 and a $\gNFW$ (left) or an Einasto (right) dark matter halo (see section~\ref{subsubsec:fit-B2-other-DM}). The black shaded region corresponds to $1\sigma$ systematic uncertainties and other areas correspond to marginal 68\% credible regions, with the mean values shown as lines.}
	\label{fig:B2-MNFW-Einasto}
\end{figure*}

As we did when analyzing the model B1, we test how robust the estimated value of $\localDM$ is to the change in shape of the dark halo. We analyze both $\rho_{\gNFW}$ and $\rho_{\rm Ein}$. The estimated quantities are presented in Tab.~\ref{tab:B2-NFW-MNFW-Einasto} and the corresponding circular velocity curves are similar to those obtained for an NFW (see Fig.~\ref{fig:B2-NFW}) and are presented in Fig.~\ref{fig:B2-MNFW-Einasto}. 

Using a $\gNFW$ distribution, with free inner slope $\gamma$, we find a consistent measurement for all the parameters with respect to the case with $\gamma = 1$ (the usual NFW profile). However, there is a noticeable increase in the error bars because of the presence of an additional free parameter (the inner slope $\gamma$).
On the other hand, an Einasto profile produces a smaller virial mass $M_{200}$ and virial radius $r_{200}$. The maximum of their posteriors are nonetheless fully compatible with the studies using an NFW or a $\gNFW$ halo.

The plot including the 2D marginal credible intervals for the parameters of Tab.~\ref{tab:B2-NFW-MNFW-Einasto} is shown in Fig.~\ref{fig:triang-rel-B2-NFW} of the appendix~\ref{appendix:triang-plots}.

Similarly to what we obtained when analyzing the baryonic model B1, the estimate of $\localDM$ from studying the model B2 is robust under the change of the dark halo, although it presents a larger variation in the maximum and credible regions of its posterior.

The main source of change in the estimate of $\localDM$ is then the different baryonic distributions considered in the model B1 (analyses of section~\ref{subsec:analysis-B1}) and B2 (analyses of this section~\ref{subsec:analysis-B2}).

\section{Discussion}
\label{sec:discussion}

In order to understand better the consistency of the results obtained in our analyses, we compare them with other estimates in the literature, both regarding $\localDM$ and other derived results. First, we comment on the local dark matter density in section~\ref{subsec:discussion-localDM}. Then we test in section~\ref{subsec:discussion-DD} the constraining power of the rotation curve method to a change in a disk structure. Since $N$-body simulations provide values for the $\cmrelation$ relation, in section~\ref{subsec:fit-cvir-Mvir-relation} we compare the 2D posterior distribution from our analyses with the $\cmrelation$ trend found in simulations. Finally, we compare the estimated mass of the Milky Way from our mass models with measurements at different distances from the Galactic centre in section~\ref{subsec:discussion-masses}.

\subsection{Estimated local dark matter density}\label{subsec:discussion-localDM}

\begin{figure}
\centering
\includegraphics[width=0.6\textwidth]{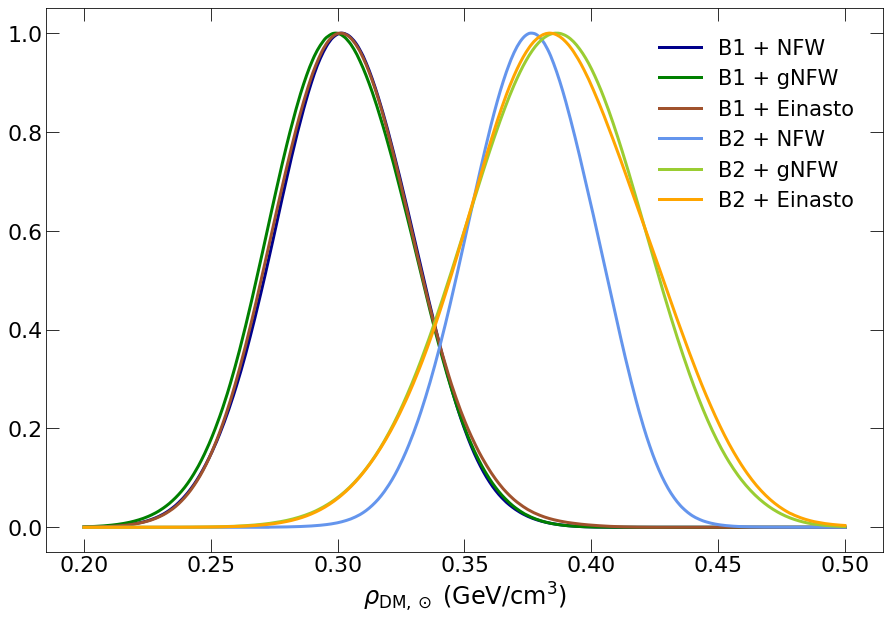}
\caption{Marginalized posterior distributions of $\localDM$ for the different cases analyzed in section~\ref{sec:analyses}, with baryonic components allowed to vary. Blue, green and brown lines correspond, respectively, to the mass models that assume an NFW, $\gNFW$ and Einasto dark halo profile. Lines showing the results of our baryonic model B1 (see section~\ref{subsec:analysis-B1}) are shown in a darker color than those associated to the analyses of the baryonic B2 model (see section~\ref{subsec:analysis-B2}), that prefer a higher $\localDM$ value.}
\label{fig:1D-localDM-all}
\end{figure}

Our study shows that the estimated dark matter parameters are affected by the choice of the distribution of baryons. 
In particular, as can be seen in Fig.~\ref{fig:1D-localDM-all} (where we present the marginalized posterior distributions of $\localDM$ for the different mass models analyzed), and in 
Tabs.~\ref{tab:Eilers-varying-othersDM} and \ref{tab:B2-NFW-MNFW-Einasto}, we find a larger value of $\localDM$ in the case of model B2 (that comprises double exponential disk profiles) than in the case of B1 (that is based on Miyamoto-Nagai disk profiles). This happens not only because of the larger mass of the B1 model, which contains around 50\% more baryonic mass than the model B2, but also because of the different shape assumed in the components, especially the disks. Assuming the same mass, the double exponential stellar disk of model B2 contributes more to the circular velocity curve at intermediate-to-large Galactocentric distances $(\text{3--10}\kpc)$ than the Miyamoto-Nagai disks of the model B1. However, the presence of two disks in the B1 model (with different scales but same mass) enhances the contribution of baryons to the circular velocity curve with respect to model B2.
Therefore, less dark matter is needed in order to fit the measured circular velocities in the case of the baryonic model B1, and the resulting $\localDM$ value is $\sim 30\%$ smaller than its estimate from the analysis of the model B2.

The difference in the two baryonic models can also be seen comparing the distance at which the dark matter contribution to the circular velocity curve becomes larger than the contribution of baryons. In particular, we obtain that for the analysis using the model B1 this distance corresponds to $R_{v_{\rm c},{\rm eq}} \sim \text{10--16}\kpc$, where the values are the minimum and maximum of the marginal 68\% posteriors for the three dark matter profiles (NFW, $\gNFW$ and Einasto).
Similarly, the dark matter mass enclosed in a sphere of radius $r$ overcomes the baryonic mass inside the sphere at a radius $r_{M,{\rm eq}} \sim \text{8--14}\kpc$. 
On the other hand, when a baryonic B2 model is used, the distances at which dark matter contributes more than baryons are $R_{v_{\rm c},{\rm eq}} \sim \text{4--11}\kpc$ and $r_{M,{\rm eq}} \sim \text{3--9}\kpc$. These distances are closer to the Galactic centre when the model B2 is used, which is consistent with a larger $\localDM$ with respect to the model B1.

Regarding a change in the dark matter profile, we find that choosing a spherical NFW, $\gNFW$ or Einasto halo does not greatly affect the $\localDM$ estimates, given a baryonic model, as can be inferred from Fig.~\ref{fig:1D-localDM-all}.

The obtained values of $\localDM$ from the analyses of the B1 and the B2 baryonic models are around $\localDM \sim 0.30\,\mathrm{GeV/cm}^3$ and $\localDM \sim 0.38\,\mathrm{GeV/cm}^3$, respectively. These values are consistent with most of the previous studies 
(some recent examples are \cite{McKee:2015hwa,Xia:2015agz,Pato:2015dua,Sivertsson:2017rkp,Karukes:2019jxv,Benito:2019ngh}), 
but smaller than others (such as the recent estimates of \cite{Buch:2018qdr,Widmark:2018ylf}).
One explanation could be the different method used to estimate $\localDM$, since \cite{Buch:2018qdr,Widmark:2018ylf} exploit the local $z$-Jeans equation method, which is based on the vertical movement of stars in a region close to the Solar System 
(see e.g. \cite{Binney:2008-book,Read:2014qva}). 
This method is likely to overestimate $\localDM$ when the assumption of equilibrium breaks down.
Observations of asymmetries in the densities and velocities of stars in the disk show that the disk is likely experiencing vertical oscillations, probably because of the passage of a massive satellite 
(see e.g. \cite{Antoja:1804.10196,Heines:1903.00607}).
In addition, the Solar System could be placed at an overdense region, compared to the surroundings, and it could also be possible that the baryonic models for gas and dust are incorrect. We further comment on this possibility in the next section.

\subsection{Sensitivity to disk uncertainties}\label{subsec:discussion-DD}

In the previous section we discussed that the main difference between the best-fit values of $\localDM$ that we obtain from the analyses of the two baryonic models, B1 and B2, arise from the difference in their disk structure. Thus, the estimate of $\localDM$ using the rotation curve method is particularly sensitive to differences in the shape of the Galactic disk. In this section we study by how much the determined value of $\localDM$ could be affected by disk uncertainties.

In principle, disk uncertainties could move $\localDM$ either up or down. However, it is rather difficult to obtain a smaller value of $\localDM$, because it would imply that the distribution of baryons decline even less rapidly, both vertically and in radius, than in the Miyamoto-Nagai disks of the model B1. Such a spread distribution seems to be incompatible with other observations~\citep{Rix:2013bi,Bland-Hawthorn:1602.07702}. On the contrary, an additional thin disk distribution would increase $\localDM$ and could explain the rather large values measured by \cite{Buch:2018qdr,Widmark:2018ylf}. Remember, however, that these authors used the $z$-Jeans equation method in order to infer $\localDM$, and it is known that local methods can be affected by disequilibria~\cite{Banik:2016yqm}.

If the overdensity found in local studies is not a feature of disequilibrium, it could be argued that the excess is produced by a dark matter disk (motivated by beyond the Standard Model physics, \cite{Read:2008fh,Purcell:2009yp,Fan:2013yva}), while another possible explanation considers that it is made of baryonic gas \cite{Widmark:2018ylf}. 
Independently of the reason for such an excess, we can assume for a moment that there really is an additional, extremely thin component in the Galaxy, that will increase $\localDM$, and study whether our analysis using the rotation curve method can determine the presence of the new component.

Therefore, we performed a simple test. We used our baryonic B1 model configuration with varying baryons plus an NFW spherical halo. To these components we added an extremely thin double exponential disk, with fixed parameters $R_d = 3.93\kpc$, $z_d = 0.038\kpc$ and $M = 7.3\times 10^{9}\msun$, that increases $\localDM$ by an amount compatible with the local observations of \cite{Buch:2018qdr,Widmark:2018ylf}.
This additional component contributes to the circular velocity curve of the Galaxy in a similar way as gas from the B2 model (see e.g. Fig.~\ref{fig:B2-NFW}). We find that the varied parameters of the fit adapt very well, acquiring values consistent with the analysis without the extra thin disk. 
Thus, the rotation curve method is unable to notice the presence of such a thin component.
However, the new component adds some extra energy density, in particular close to the Galactic plane.
If we interpret the energy density of the new thin disk as a dark, non visible source, we should add an extra $\Delta \localDM = \{4.7, 2.4, 0.3\}\GeVcm$ for a distance from the Galactic plane of $z = \{ 0,25,100 \}\,\mathrm{pc}$. Since the Earth is at around $z_\odot \approx 25\,\mathrm{pc}$ from the plane \citep{Bland-Hawthorn:1602.07702}, this would imply a local dark matter density 8 times larger than the one obtained in the same analysis without the extra thin component.

We insist that we are not arguing in favor of the existence of the additional thin disk. But it is important to show that the rotation curve method is insensitive to the existence of such a component, and the obtained values of $\localDM$ from our analyses might be smaller than the real $\localDM$ value if there was indeed a new component whose distribution is too spread out to be noticed from the analysis of the Milky Way's rotation curve.

\subsection{Studying the $\cmrelation$ relation}\label{subsec:fit-cvir-Mvir-relation}

In order to investigate further the consistency with other studies, here we explore the relation between the two main virial quantities, $c_{200}$ and $M_{200}$. Note, however, that the relation that we want to study in this section is independent of the correlation found in our analyses between the two fitted dark matter quantities; such correlation between $c_{200}$ and $M_{200}$, depicted in Fig.~\ref{fig:MvirCvir-rep}, is a result of the way that both parameters affect the fit of the dark matter distribution to the data. Given the values of $v_{\rm c}$ determined in \cite{Eilers:1810.09466} and for a fixed baryonic mass distribution, a more concentrated dark halo (with larger $c_{200}$) needs to have a smaller virial mass $M_{200}$ in order to fit the data, as shown in Fig.~\ref{fig:MvirCvir-rep}. However, what we are going to study in this section is the comparison of the relation between the two virial parameters, $c_{200}$ and $M_{200}$, found in our analyses with respect to the same relation obtained in $N$-body simulations 
(see for instance \cite{Dutton:2014xda,Klypin:2014kpa,Bose:2019hwt}).

Note that in our analyses we have used flat priors on the values of the two main virial quantities, $c_{200}$ and $M_{200}$. However, $N$-body simulations reveal a characteristic relation between these two quantities (which we reiterate is different than the relation shown in Fig~\ref{fig:MvirCvir-rep}), that can be expressed as
\begin{equation}\label{eq:cMrel}
\log_{10} c_{200} = a + b \log_{10} \left( M_{200}/[10^{12} h^{-1} \msun] \right),
\end{equation}
where $a = 0.520$, $b = -0.101$ for an NFW halo, and $a = 0.459$, $b = -0.130$ for an Einasto profile \citep{Dutton:2014xda}.

\begin{figure*}
    \centering
    \includegraphics[width=0.43\textwidth]{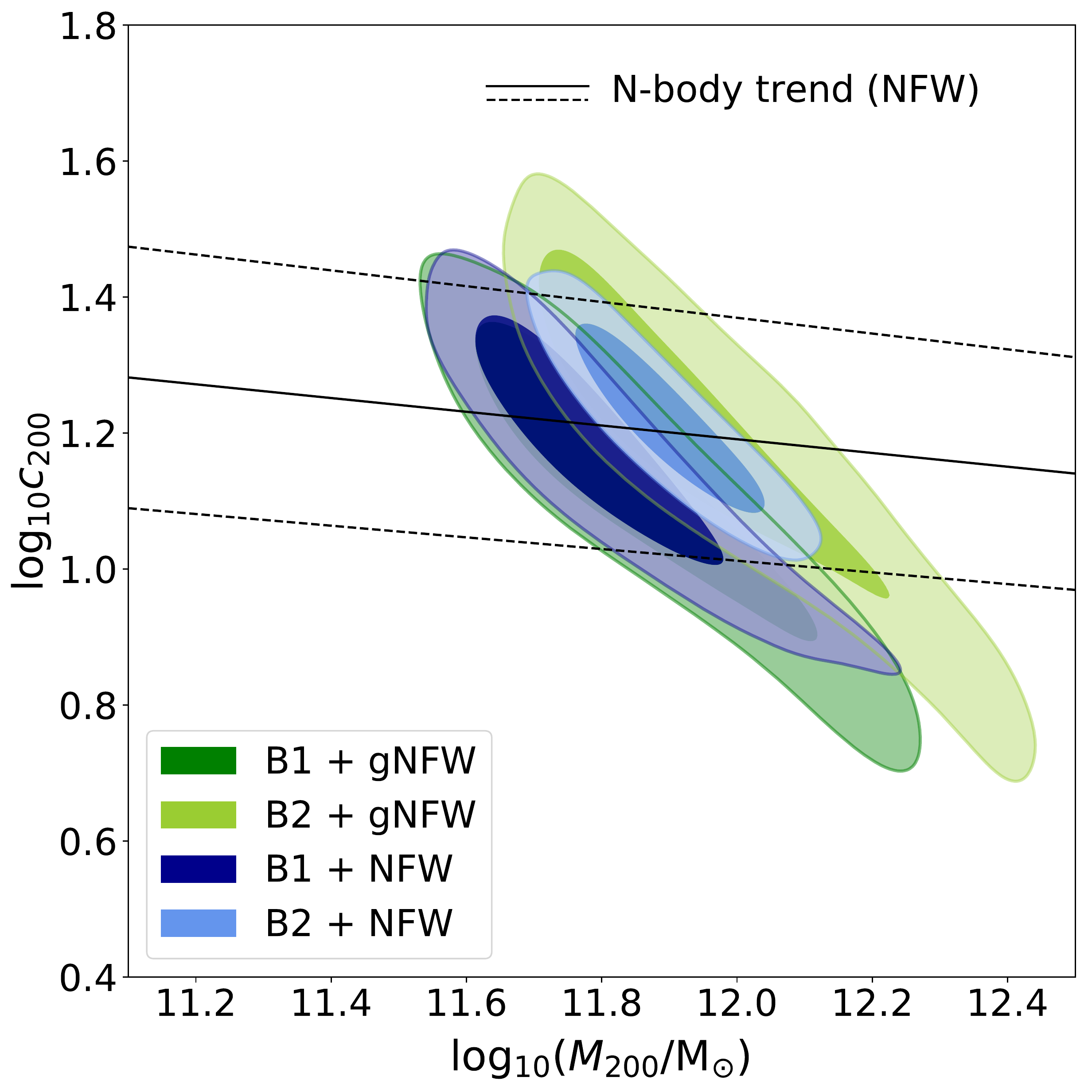}
    \includegraphics[width=0.43\textwidth]{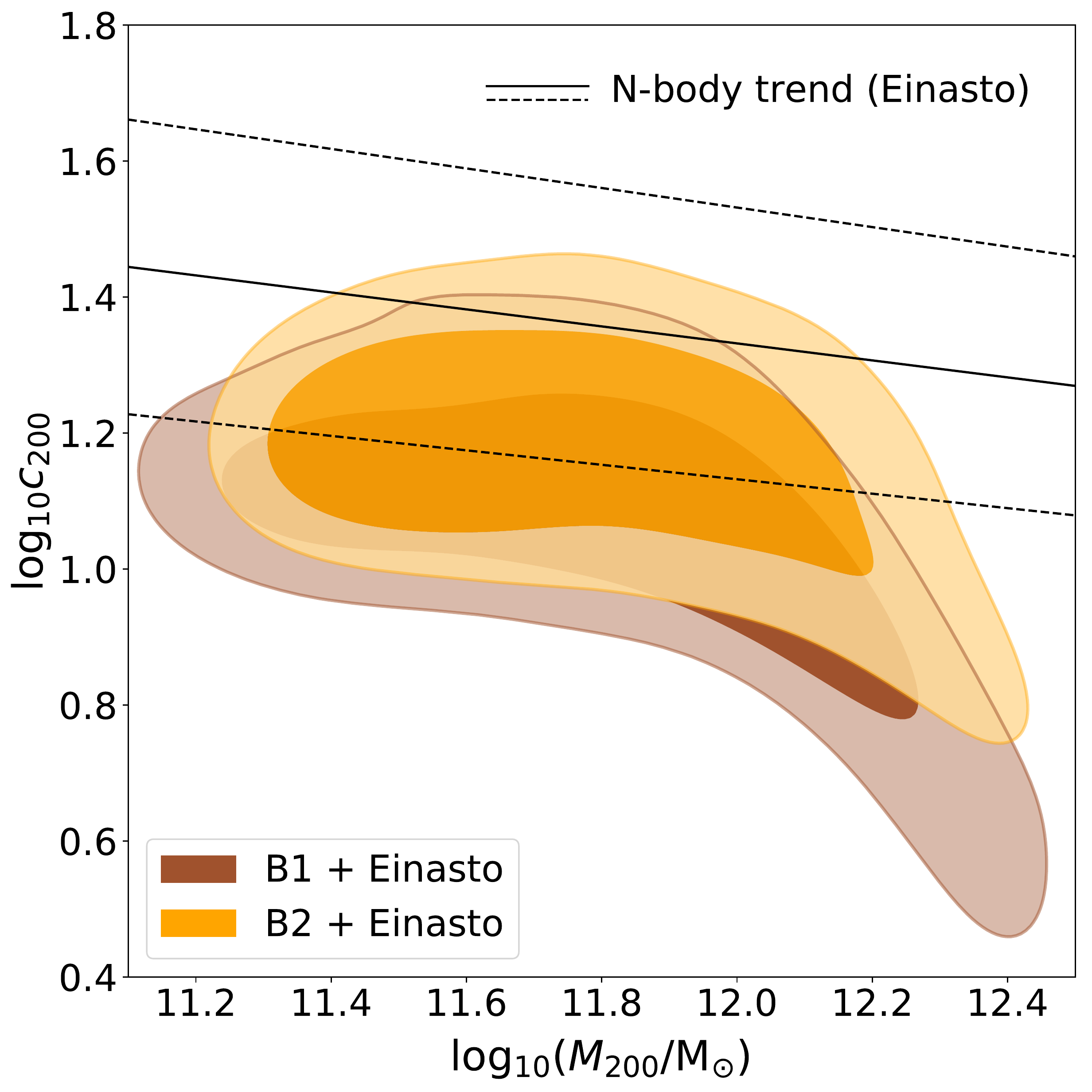}
    \caption{Marginal 68\% and 95\% credible regions for the analyses including an NFW or $\gNFW$ (left panel), or an Einasto (right panel) dark halo. The $\cmrelation$ relation obtained from $N$-body simulations \protect\citep{Dutton:2014xda} is also shown, with dashed lines corresponding to its estimated $1\sigma$ uncertainty from \protect\cite{Udrescu:2018hvl}.}
    \label{fig:logCv-logMv}
\end{figure*}

In Fig.~\ref{fig:logCv-logMv} we show the comparison between our analyses and $N$-body simulations. In the left panel the fitted dark matter halo is an NFW (or a $\gNFW$) and in the right panel we show the same when an Einasto profile is considered.

All marginal credible regions shown in the left panel are perfectly centered at the $N$-body trend, which assumes an NFW shape of the halo. The studies of the baryonic model B1 prefer a slightly less massive halo and the 2D posteriors of our analyses are wider for the $\gNFW$ profile since it has an extra free parameter.

For those cases that include an Einasto profile the comparison is less attractive, but the result of our analyses and the $N$-body trend are still fully compatible. As shown in Tabs.~\ref{tab:Eilers-varying-othersDM} and \ref{tab:B2-NFW-MNFW-Einasto}, the preferred $M_{200}$ value is smaller for an Einasto halo than for a $\gNFW$. However, an Einasto profile fits more loosely to data and the 68\% and 95\% posteriors extend beyond those found with a $\gNFW$, except for the largest allowed values of $c_{200}$ (see Fig.~\ref{fig:logCv-logMv}).

Overall, our results respect what is found in $N$-body simulations, indicating that our Galactic model behaves as expected.
However, it is worth mentioning that the $\cmrelation$ relations of Eq.~\eqref{eq:cMrel} were obtained from dark matter only simulations, and the lines in Fig.~\ref{fig:logCv-logMv} from \cite{Dutton:2014xda} could change if baryons are included 
(see e.g. \cite{Sawala:2015cdf,Kelso:2016qqj}).

\subsection{Estimated Galactic mass}\label{subsec:discussion-masses}

Another quantity that we can compare with observations is the total dynamical mass of the Galaxy at different distances from the Galactic centre.
In Tab.~\ref{tab:checks-with-literature} we present the values obtained in our main studies, with varying baryonic parameters and flat priors on the virial quantities, and compare them with a selection of observational estimates. As can be seen in the table, our results are compatible with other studies in a wide range of values; however, since the reach of the data that we have used only covers up to $\sim 25\kpc$, the error of our estimated masses grow at larger distances, making our estimate at the largest distance of $300\kpc$ comparatively smaller than the value found by \cite{Watkins:2010fe}.

This difference is another indication that the constraining power of circular velocities is limited.
 Although current well measured $v_{\rm c}$ data have increased their reach up to $25\kpc$, this distance is not enough to put robust constraints on the virial mass and concentration parameters of the halo, since its real shape, in particular beyond $25\kpc$, could be very different from the assumed profiles of this work. 
However, we decided to present the estimated Galactic mass up to such large distances to facilitate the comparison of the properties of our fitted mass models with those that can be found in the literature.

\begin{table*}
    \centering
    \begin{tabular}{c||c|c|c||c|c|c||c|l}
    & \multicolumn{3}{c||}{Baryonic model B1} & \multicolumn{3}{c||}{Baryonic model B2} \\
    \hline
         & NFW & $\gNFW$ & Einasto & NFW & $\gNFW$ & Einasto & Reference value & Ref. \\
         \hline
    $M_{\mathrm{MW}}(R<300\,\mathrm{kpc})\;[10^{11}\,\mathrm{M}_\odot]$ & 6.5--9.8 & 6.5--10.9 & 3.6--15.6 & 7.9--11.9 & 6.8--12.5 & 3.4--10.6 & $14\pm 3$ & \cite{Watkins:2010fe} \\
    \hline
    $M_{\mathrm{MW}}(R<100\,\mathrm{kpc})\;[10^{11}\,\mathrm{M}_\odot]$ & 4.4--5.9 & 4.5--6.4 & 3.6--7.8 & 5.1--6.9 & 4.8--7.2 & 3.4--7.2 & $4.1\pm 0.4$ & \cite{Gibbons:2014ewa} \\
    \hline
    $M_{\mathrm{MW}}(R<50\,\mathrm{kpc})\;[10^{11}\,\mathrm{M}_\odot]$ & 3.2--3.9 & 3.3--4.0 & 3.1--4.5 & 3.6--4.4 & 3.5--4.5 & 3.2--4.6 & $3.7^{+0.4}_{-0.3}$ & \cite{Eadie:1810.10036}\\
    \hline
    $M_{\mathrm{MW}}(R\lesssim 20\,\mathrm{kpc})\;[10^{11}\,\mathrm{M}_\odot]$ & 1.9--2.0 & 1.9--2.0 & 1.8--2.0 & 1.9--2.1 & 1.9--2.1 & 1.9--2.1 & $2.1^{+0.4}_{-0.3}$ & \cite{Watkins:1804.11348} \\
    \hline
    \end{tabular}
    \caption{Comparison between the mass of the Milky Way obtained from the mass models of our analyses and estimates from other studies. Marginal 68\% credible regions are shown. The first three columns correspond to our baryonic model B1 and columns 4--6 to the baryonic model B2. Different radial distances are considered. 
    }
    \label{tab:checks-with-literature}
\end{table*}

\section{Conclusions}\label{sec:conclusions}

Our goal in this work was both to estimate the value of $\localDM$ from the rotation curve of the Galaxy and to study the robustness of the determination. This last point was addressed using different Galactic mass models, with different dark matter and baryonic density distributions. In order to estimate $\localDM$, we fitted the parameters of the mass models to the precise measurements (using Gaia DR2 data) of the Milky Way's circular velocity presented in \cite{Eilers:1810.09466}.

We examined two baryonic models, referred to in the text as B1 and B2 (see section~\ref{sec:models} for the description of the models), and tested three different dark matter spherical halos (NFW, $\gNFW$ and Einasto). The main difference between the baryonic models stands in the fact that B1 is denser than the model B2 in the intermediate Galactic radii regions and out of the Galactic plane.

Given a baryonic mass model, we found the value of $\localDM$ to be robust under the change of the spherical dark matter halo. However, $\localDM$ was more sensitive to a change in the baryonic model.
In particular, in the analyses of the model B1 we found a value of $\localDM \simeq 0.30\GeVcm$ (see Tab.~\ref{tab:Eilers-varying-othersDM}), while $\localDM \simeq 0.38\GeVcm$ was found instead for the baryonic model B2 (see Tab.~\ref{tab:B2-NFW-MNFW-Einasto}). 
Notice that the $68\%$ region for the baryonic model B1 goes up to $\localDM \approx 0.33\GeVcm$, while the $68\%$  region for the B2 model goes down to $\localDM \approx 0.35 \GeVcm$.  One can see that the regions do not overlap at the $68\%$ level, but they do at $95\%$.
Thus the baryonic models are not totally incompatible. 

Although the difference between models B1 and B2 is not extremely large, it is important, in particular given that usually people only cite the error bars for one baryonic model.
Within one single model the largest $1 \sigma$ error bars on $\localDM$ correspond to $\pm 0.036 \GeVcm$ (for the case of baryonic model B2 and Einasto dark matter profile), see Tab.~\ref{tab:B2-NFW-MNFW-Einasto}.  On the other hand, when we take into account the variety of mass models considered in this paper, in particular the uncertainty in the baryonic profile by considering both models B1 and B2, the $1 \sigma$ uncertainty range reaches  $0.149 \GeVcm$.  This error is at least twice that from any of the individual mass models.

We also tested the robustness of the rotation curve method to estimate $\localDM$ assuming the existence of an hypothetical thin disk, together with an NFW dark halo and the baryonic components of the B1 model. In this scenario the fit to the $v_{\rm c}$ data is not able to perfectly distinguish the presence of such thin disk, since it barely contributes to the Galactic circular velocity curve. 
Thus, uncertainties in the Galactic disk can significantly increase the uncertainty in the estimate of $\localDM$,
that could be more than a factor 8 larger if an additional thin disk is present in the Milky Way.

Furthermore, we compared the resulting virial quantities of our analyses with the $\cmrelation$ trend from $N$-body simulations, finding them to be consistent.
We additionally compared the total Galactic mass resulting from our analyses with respect to observational estimates at different distances from the Galactic centre. 
This comparison shows that the constraining power of circular velocities is limited. The reach of current $v_{\rm c}$ measurements have extended its precision up to $25\kpc$, but this distance is still not enough to put robust constraints on the parameters (virial mass and concentration) of the halo. Indeed the halo shape could be different from the assumed profiles of this work, in particular at larger distances from the Galactic centre.

Overall, we proved that the estimate of $\localDM$ from Milky Way's rotation curve measurements depends on the assumed shape for the mass distribution, regardless of the precision of the $v_{\rm c}$ measurements.

In order to constrain $\localDM$ in a more robust way, more data apart from circular velocities need to be taken into account, like phase-space information of different type of tracers such as halo stars that travel far from the disk plane. We plan to do that in future works, also combining different methods to that of the rotation curve (for instance we can explore a combination with local methods, such as the vertical Jeans equation method) in a search for an estimate of $\localDM$ as precise and consistent as possible.

\begin{figure}[t]
\centering
\includegraphics[width=\textwidth]{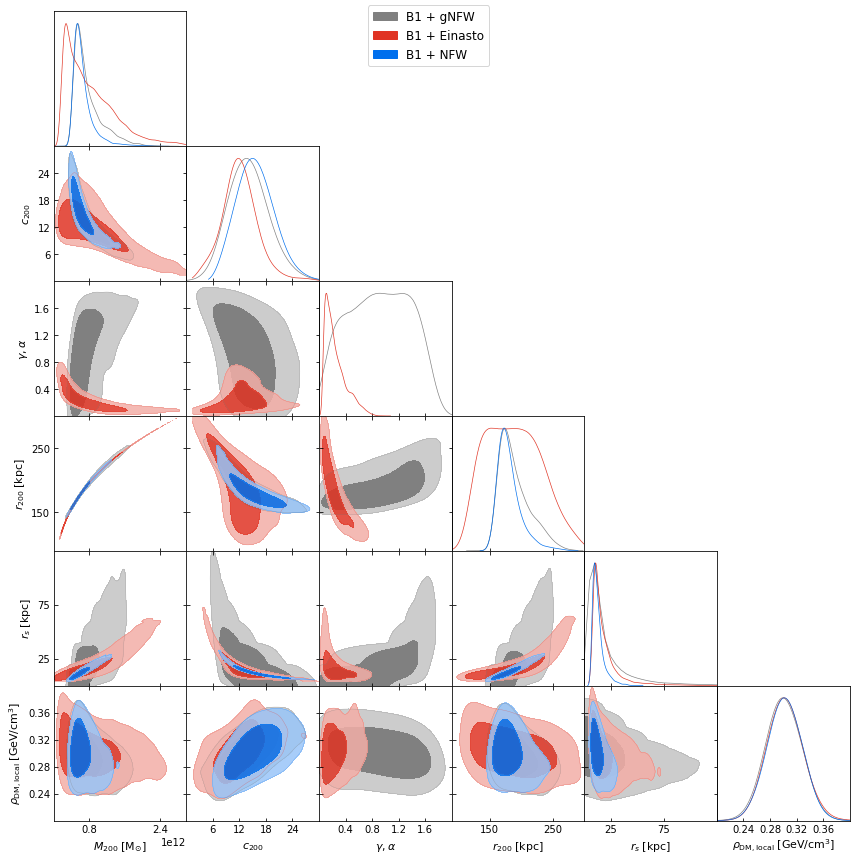}
\caption{Marginal 68\% and 95\% credible regions for the parameters presented in Tab.~\ref{tab:Eilers-varying-othersDM}, corresponding to the analyses of section~\ref{subsec:analysis-B1} that include the baryonic model B1. 
We note that the parameters $\gamma$ and $\alpha$ apply to different dark matter models (gNFW and Einasto) which is why the regions for the different models are not necessarily expected to overlap.}
\label{fig:triang-rel-B1-NFW}
\end{figure}

\begin{acknowledgments}

PFdS thanks Justin Alsing for useful discussions about our Bayesian analyses, and Sofia Sivertsson, Eric F. Bell and Oleg Gnedin for their comments and discussions.
PFdS, KM and KF acknowledge support by the Vetenskapsr{\aa}det (Swedish Research Council) through contract No. 638-2013-8993 and the Oskar Klein Centre for Cosmoparticle Physics.
KF acknowledges support from DoE grant DE-SC007859 and the LCTP at the University of Michigan.
MV and KH are supported by NASA-ATP award NNX15AK79G to the University of Michigan.
PFdS thanks the LCTP at the University of Michigan for the hospitality received while this work was been finalized.

\end{acknowledgments}


\appendix
\section{Triangular plots of the dark matter parameters}
\label{appendix:triang-plots}

\begin{figure}[H]
\centering
\includegraphics[width=\textwidth]{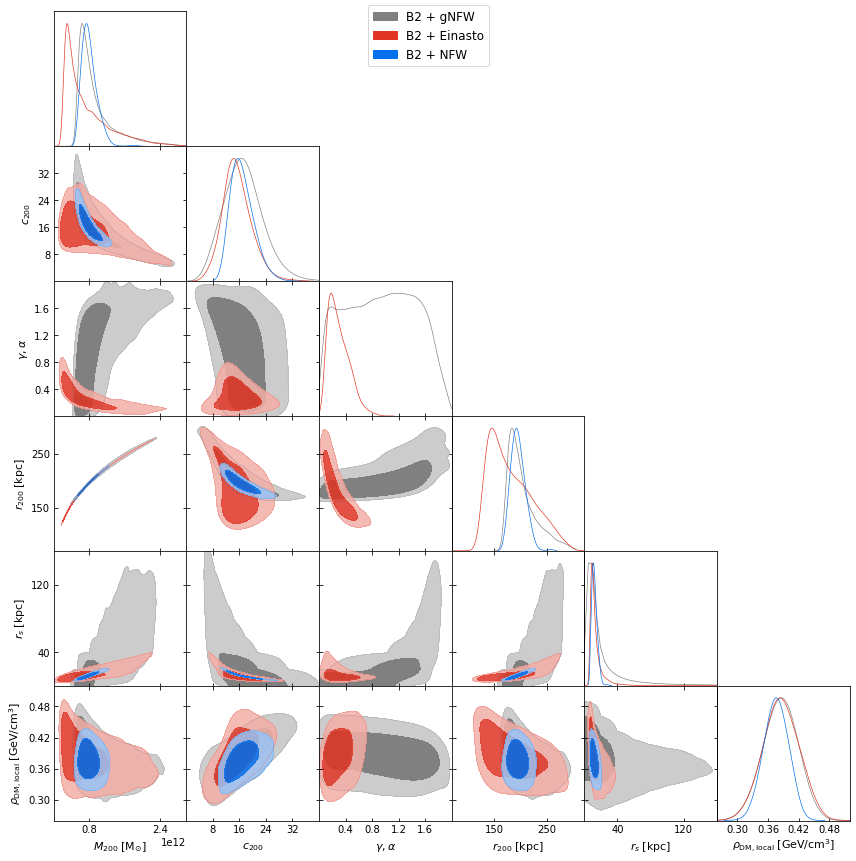}
\caption{Marginal 68\% and 95\% credible regions for the parameters presented in Tab.~\ref{tab:B2-NFW-MNFW-Einasto}, corresponding to the analyses of section~\ref{subsec:analysis-B2} that include the baryonic model B2.
 Again, the parameters $\gamma$ and $\alpha$ apply to different dark matter models (gNFW and Einasto) which is why the regions for the different models are not necessarily expected to overlap.}
\label{fig:triang-rel-B2-NFW}
\end{figure}

In this appendix we present the triangular plots corresponding to the 2D marginal 68\% and 95\% credible regions of the dark matter fitted and derived parameters of our analyses. Figure~\ref{fig:triang-rel-B1-NFW} shows the triangular plot for the analyses including the baryonic model B1 (section~\ref{subsec:analysis-B1}). Figure~\ref{fig:triang-rel-B2-NFW} shows the triangular plot for the analyses including the baryonic model B2 (section~\ref{subsec:analysis-B2}).

\bibliographystyle{JHEP_inspired}


\end{document}